# Bright and Efficient Perovskite Light Emitting Electrochemical Cells Leveraging Ionic Additives

*Masoud Alahbakhshi[1], Aditya Mishra[2], Ross Haroldson[3], Arthur Ishteev[6], Jiyoung Moon[2], Qing Gu[1], Jason D. Slinker[2,3]\* and Anvar A. Zakhidov[3,4,5]\**

[1]Department of Electrical and Computer Engineering, The University of Texas at Dallas, 800 West Campbell Road, Richardson, Texas 75080-3021, United States.

[2]Department of Materials Science and Engineering, The University of Texas at Dallas, 800 West Campbell Road, Richardson, Texas 75080-3021, United States.

[3]Department of Physics, The University of Texas at Dallas, 800 West Campbell Road, Richardson, Texas 75080-3021, United States.

[4]NanoTech Institute, The University of Texas at Dallas, 800 West Campbell Road, Richardson, Texas 75080-3021, United States.

[5]Department of Nanophotonics and Metamaterials, ITMO University, St. Petersburg, Moscow, Russia.

[6]Laboratory of Advanced Solar Energy, NUST MISiS, Moscow, 119049, Russia

AUTHOR INFORMATION

**Corresponding Author**

*slinker@utdallas.edu, *zakhidov@utdallas.edu

ABSTRACT: Perovskite light emitting diodes (PeLEDs) have drawn considerable attention for their favorable optoelectronic properties. Perovskite light emitting electrochemical cells (PeLECs) – devices that utilize mobile ions – have recently been reported but have yet to reach the performance of the best PeLEDs. We leveraged a poly(ethylene oxide) electrolyte and lithium dopant in $CsPbBr_3$ thin films to produce PeLECs of improved brightness and efficiency. In particular, we found that a single layer PeLEC from $CsPbBr_3$:PEO:$LiPF_6$ with 0.5% wt. $LiPF_6$ produced highly efficient (22 cd/A) and bright (~15000 cd/m$^2$) electroluminescence. To understand this improved performance among PeLECs, we characterized these perovskite thin films with photoluminescence (PL) spectroscopy, scanning electron microscopy (SEM), atomic force microscopy (AFM), X-ray photoelectron spectroscopy (XPS), and X-ray diffraction (XRD). These studies revealed that this optimal $LiPF_6$ concentration improves electrical double layer formation, reduces the occurrence of voids, charge traps, and pinholes, and increases grain size and packing density.

**TOC GRAPHICS**

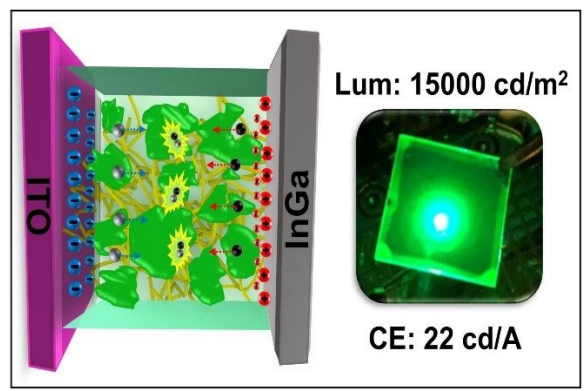

Perovskite light-emitting diodes (PeLEDs) based on inorgano−organometallic halide perovskites, such as $CH_3NH_3PbX_3$ and $CsPbX_3$ (X = Cl, Br, or I), have attracted much attention due to their low-temperature solution processability, high color purity with narrow spectral width (FWHM of 20 nm), band gap tunability and large charge carrier mobility.[1-4] To date, devices based on these perovskites have achieved high luminance in excess of 10000 cd/m² with high efficiencies (EQE ~10%), comparable to organic LEDs and quantum dot (QD) LEDs.[1-7]

Interestingly, effects such as hysteresis and high capacitance in perovskite semiconductor devices suggest that ion motion can largely influence device operation. In this vein, researchers have recently been investigating perovskite materials in light-emitting electrochemical cell (LEC) architectures instead of traditional LEDs.[8-11] These LEC devices (PeLEC leverage ion redistribution to achieve balanced and high carrier injection, resulting in high electroluminescence efficiency. Due to this mechanism, LEC devices can be prepared from a simple architecture consisting of a single semiconducting composite layer sandwiched between two electrodes. In addition, they can operate at low voltages below the bandgap, yielding highly efficient devices. Recently, perovskite LECs (PeLECs) have been reported and show promise as electroluminescent devices.[8-11] However, these PeLECs are generally limited to luminance maxima of 1000 cd/m² or lower, below what has been typically observed in PeLEDs. This disparity suggests that further understanding and refinement of PeLEC materials and devices could produce significant improvements of brightness, efficiency, and other performance metrics. To this end, we fabricated a highly efficient (22 cd/A) and bright (~15000 cd/m²) single layer LEC based on a cesium lead halide perovskite, $CsPbBr_3$. To achieve high performance, a polyelectrolyte and additive mobile ions were carefully selected to achieve optimal ionic redistribution and doping effects. To understand the nature of this performance and its correlation with materials properties, we characterize these perovskite thin films and devices with photoluminescence spectroscopy, scanning electron microscopy (SEM), atomic force microscopy (AFM), X-ray photoelectron spectroscopy (XPS), and X-ray diffraction (XRD).

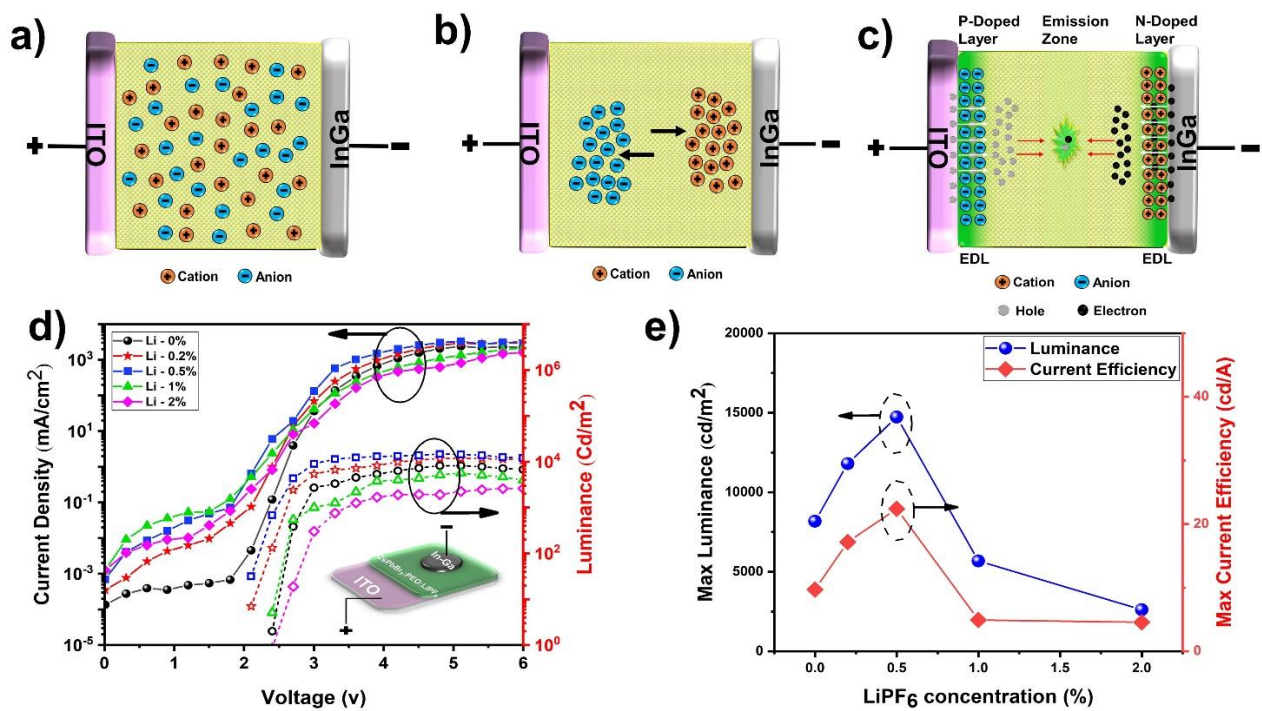

Figure 1. Illustration of PeLEC device operation and device characteristics. a) Initial LEC state showing ions uniformly distributed throughout the active layer. b) Intermediate LEC state showing cations drifting toward the cathode and anions toward the anode. c) Steady-state LEC operation with ions accumulated at the electrodes and light emission upon current injection. d) Current density and luminance versus voltage for In-Ga/CsPbBr$_3$:PEO:LiPF$_6$ LEC/ITO devices. The inset shows the layering of In-Ga/CsPbBr$_3$:PEO:LiPF$_6$/ITO. e) Maximum luminance and current efficiency as a function of LiPF$_6$ concentration.

We first consider the distinct stages of LEC device operation, which are illustrated in Figures 1a-c. Initially, ions are uniformly distributed in the film (Figure 1a). In response to an applied bias, cations drift toward and accumulate near the cathode, and anions likewise move toward and pack near the anode (Figure 1b). This leads to an electric double layer (EDL) formation at each electrode that induces higher electric fields, decreased width of the potential barriers (doping), and enhanced injection of electrons and holes that is insensitive to the workfunction of the electrodes (Figure 1c).[12-15] These injected carriers are transported through the bulk and radiatively recombine in the center of the device. The key features needed for successful LEC operation are therefore: 1) A sufficient concentration of mobile anions and cations; 2) Efficient transport of ions through the bulk for balanced EDL formation at the anode and cathode, leading to efficient charge injection; 3) Facile transport of electrons and holes through the semiconductor (which, for our system, requires a percolating network of the perovskite); 4) Efficient light emission upon recombination of the electrons and holes in the bulk, typically supported by a high quantum yield of the film. In our specific case, high luminescence efficiency is supported by the spectral properties of the $CsPbBr_3$ perovskite. To satisfy the other requirements, we introduce $LiPF_6$ salt, a salt that we have previously used to attain high performance in LECs utilizing ionic transition metal complexes.[16-18] We prepared films with an optimal concentration of the polymer electrolyte poly(ethylene oxide) (PEO) and systematically studied the effect of $LiPF_6$ salt addition.

The LECs were constructed with a single layer of spin cast perovskite film using an ITO anode, a $CsPbBr_3$:PEO:$LiPF_6$ perovskite composite (1:0.8 weight ratio $CsPbBr_3$:PEO, various Li weight fractions), and an In-Ga eutectic cathode, as illustrated in the inset of Figure 1d (see Supporting Information Figure S1 and text for details of fabrication and testing). Also in Figure 1d, the current density versus voltage (J *vs* V) and luminance versus voltage (L *vs* V) graph of our devices with different ratios of LiPF$_6$ is shown. Figure 1e presents the maximum luminance and current efficiency from the data of Figure 1d. All of the devices showed green electroluminescence centered near 528 nm with a FWHM of 20 nm (See Supporting information Figure S2). The reference device (CsPbBr$_3$:PEO with no LiPF$_6$) shows a turn-on voltage (V$_{on}$) of 2.5 V and a substantial maximum luminance of 8175 cd/m$^2$ at 5.5 V, with a maximum current efficiency of 9.0 cd/A. This substantial luminance for a PeLEC is accounted for the optimized PEO electrolyte concentration within the film. As LiPF$_6$ is added into the CsPbBr$_3$:PEO film, the luminance and efficiency maxima increase gradually, peak at an optimal concentration, and then decrease. The best performance was achieved with a 0.5% LiPF$_6$ weight percentage, which showed a V$_{on}$ of 1.9 V (below the bandgap), a maximum luminance of 14730 cd/m$^2$ at 5.4 V, and a maximum current efficiency of 22.4 cd/A. Notably, the current efficiency (Figure 1e) was enhanced ~2.5X compared to the pristine CsPbBr$_3$:PEO LEC. To the best of our knowledge, the highest values of luminance and current efficiency that we achieved are the best reported values so far for perovskite LECs (as distinct from LEDs). Full performance metrics for all device formulations are noted in Supporting Information Table S1, and comparisons to other literature efforts are reported in Supporting Information Table S3.

Further investigation of the devices reveals other key features. Notably, all concentrations of LiPF$_6$ enhanced current density at low operating voltage (Figure 1d), indicating LiPF$_6$ initially assists with monopolar charge injection presumably through EDL formation and p- and n-doping effects. As voltage is increased (Figure 1d), turn on voltage is generally lowered by LiPF$_6$ addition, denoting its ability to assist in bipolar injection. Luminance and current efficiency are enhanced by adding 0.2% or 0.5% LiPF$_6$, but diminished by higher concentrations (Figure 1d), suggesting competing processes affect device performance. We also observe that the optimal 0.5% LiPF$_6$ concentration considerably reduces the hysteresis in current density curves from voltage cycling (Supporting Information Figure S3). We also measured the stability of devices under constant voltage operation and found that 0.5% LiPF$_6$ can improve the stability threefold in comparison with control devices (Supporting Information Figure S4), a favorable result in view of previous reports.[19] We performed additional study to gain further mechanistic and phenomenological understanding of these observations.

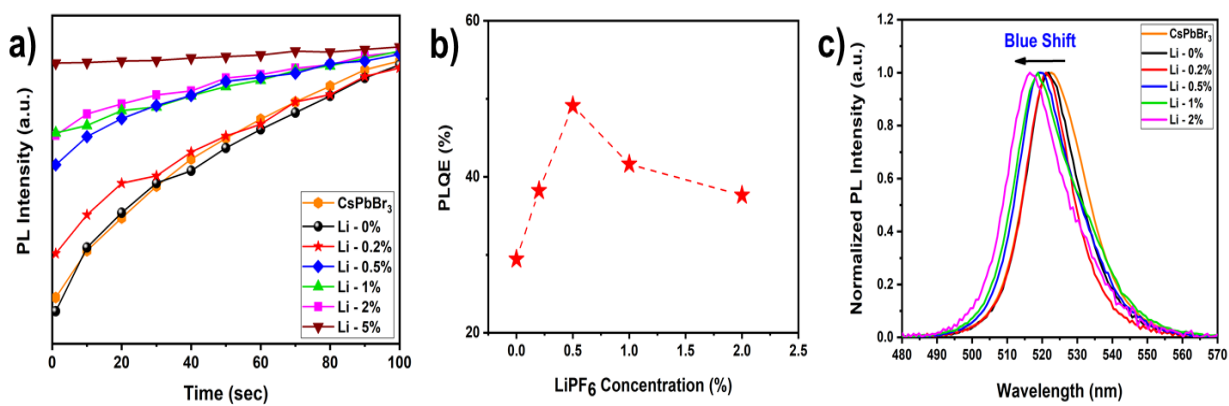

Figure 2. Photoluminescence properties of thin films of CsPbBr$_3$ and CsPbBr$_3$:PEO:LiPF$_6$. a) Photoluminescence intensity as a function of time ($\lambda_{ex}$= 405 nm). b) Photoluminescence quantum efficiency (PLQE) versus LiPF$_6$ concentration for CsPbBr$_3$:PEO:LiPF$_6$ films. c) Photoluminescence spectra of CsPbBr$_3$ and CsPbBr$_3$:PEO:LiPF$_6$.

For further investigations of the optical, electronic, and morphological states of our perovskite films, we prepared CsPbBr$_3$:PEO:LiPF$_6$ thin films through an optimized one stage spin-coating process and vacuum treatment before annealing (Supporting Information Figure S5 and text). We subsequently measured the photoluminescence (PL) intensity versus time of these films (Figure 2a). For CsPbBr$_3$ and CsPbBr$_3$:PEO films, the PL intensity dramatically increases with time. As LiPF$_6$ is added to CsPbBr$_3$:PEO films, the PL dynamic trends towards constant intensity. This PL trend with lithium doping can be understood from reduced trapping.[20] In the absence of lithium dopants, unfilled trap states are present, potentially due to grain boundaries, vacancies, and other imperfections that create nonradiative decay states in the middle of the optical gap. These trap states must first be filled with electronic carriers before steady PL can be achieved. Lithium salt addition produces filled trap states, consistent with doping concepts (see Supporting Information Figure S6), since $Li^+$ promotes n-doping by electrons that fill the trap. The morphological studies described below can further clarify the nature of trap suppression.

To understand the influence of $LiPF_6$ doping on steady-state quantum yield, we measured the absolute photoluminescence quantum efficiency (PLQE) for $CsPbBr_3$:PEO:$LiPF_6$ films with an integrating sphere according to the methods described by de Mello *et al*. and Porrès *et al*.[21-22] Figure 2b reveals that the PLQE increases from 29% for a $CsPbBr_3$:PEO film to a maximum of 49% for a 0.5% $LiPF_6$ film of $CsPbBr_3$:PEO:$LiPF_6$, and then is lowered as the $LiPF_6$ concentration is further increased. This PLQE concentration trend of $CsPbBr_3$:PEO:$LiPF_6$ films qualitatively follows the concentration dependences of luminance and current efficiency for PeLEC devices (Figure 1e). This reinforces our assertion that $LiPF_6$ suppresses nonradiative trapping states to improve emission yield in thin films and devices.

PL spectra were also measured to ascertain the impact of lithium doping on optical emission from $CsPbBr_3$:PEO:$LiPF_6$ thin films (Figure 2c). The PL spectra show a gradual blue shift from $\lambda_{max}$ = 522 nm for pure $CsPbBr_3$ thin film to $\lambda_{max}$ = 516 nm for the $CsPbBr_3$:PEO:$LiPF_6$ (5%) thin film. This blue shift may be attributed to the Burstein-Moss effect (Supporting Information Figure S6c). According to this theory, electrons populate states in the conduction band due to doping, promoting the Fermi level higher in the conduction band. Electrons from the top of the valence band can only be excited into empty conduction states above the Fermi level, consistent with the

Pauli exclusion principle.[23-24] Similar results have been reported for Sb and Bi doping in perovskites.[25-26]

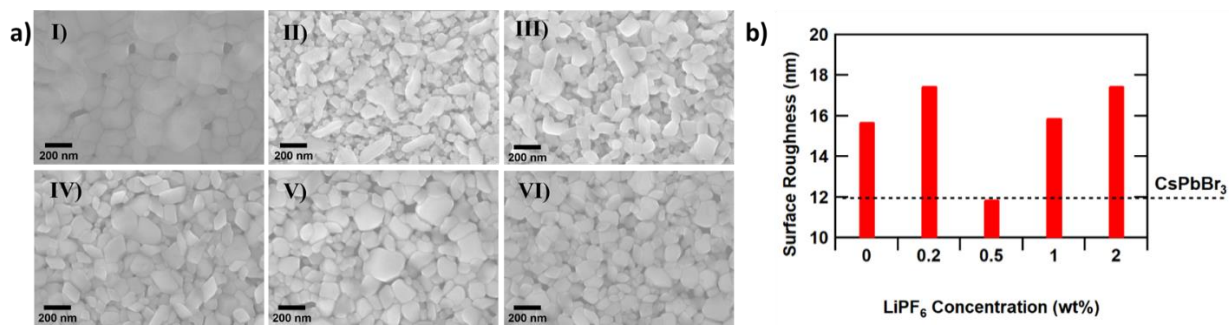

Figure 3. Morphological study of $CsPbBr_3$:PEO:$LiPF_6$ thin films by SEM and AFM. a) SEM for thin films of I) $CsPbBr_3$, II) $CsPbBr_3$:PEO, III) $CsPbBr_3$:PEO:$LiPF_6$ (0.2%), IV) $CsPbBr_3$:PEO:$LiPF_6$ (0.5%), V) $CsPbBr_3$:PEO:$LiPF_6$ (1%), and VI) $CsPbBr_3$:PEO:$LiPF_6$ (2%). b) Surface roughness of $CsPbBr_3$:PEO:$LiPF_6$ thin films versus $LiPF_6$ concentration as measured by AFM. The horizontal dotted line denotes the surface roughness of the pristine $CsPbBr_3$ film.

To further understand the morphological influence of the beneficial electrical and optical properties afforded by electrolyte and lithium addition, we studied thin perovskite films by scanning electron microscopy (SEM) and atomic force microscopy (AFM). The SEM image of the pristine $CsPbBr_3$ film (Figure 3a,I) reveals randomly oriented grains and substantial pinholes in the film. Addition of PEO (Figure 3a,II) reduces the average grain and pinhole sizes, but significant pinholes still remain. Subsequent addition of low concentrations ($\leq 2\%$) of $LiPF_6$ in $CsPbBr_3$:PEO increases the average grain size (Supporting Information Table S2) and decreases the number and size of pinholes (Figure 3a, III-VI, Supporting Information Figure S7). The largest and most monodispersed grain sizes are observed for 0.5% and 1.0% $LiPF_6$ concentrations. Also, the 0.5% $LiPF_6$ dopant film (Figure 3a, IV) provides the most continuous network of perovskite crystals with the lowest amount of pinholes. This correlates well with the high luminance and efficiency of the 0.5% LiPF$_6$ doped LEC device (Figure 1e): reducing the number of pinholes limits detrimental leakage current, and the interpenetrating network of perovskite crystals supports facile transport of electrons and holes for efficient electroluminescence. On the other hand, excess LiPF$_6$ (5%) was detrimental to the quality of the thin film, leading to discontinuous films (Supporting Information Figure S8). This is likely due to the formation of lithium dendrites and subsequent phase separation.[27]

In Figure 3b, we relate the average surface roughness of the CsPbBr$_3$:PEO:LiPF$_6$ films with various percentages of LiPF$_6$ as measured by AFM (Supporting Information Figure S9). We observe a reduction in the root-mean-square roughness from 15.7 nm for the pristine CsPbBr$_3$:PEO film to 11.9 nm for the 0.5% LiPF$_6$ doped film, and an increase in surface roughness for higher concentrations. Again, this correlates with the optimal LiPF$_6$ device concentration, as low surface roughness improves the spatial uniformity of the electroluminescence and limits pinhole formation. Hence, the overall interpretation from SEM and AFM analysis indicates that 0.5% LiPF$_6$ corresponds to the optimal concentration for smooth films with large, percolating perovskite grains and minimal pinholes, all beneficial for superior device performance.

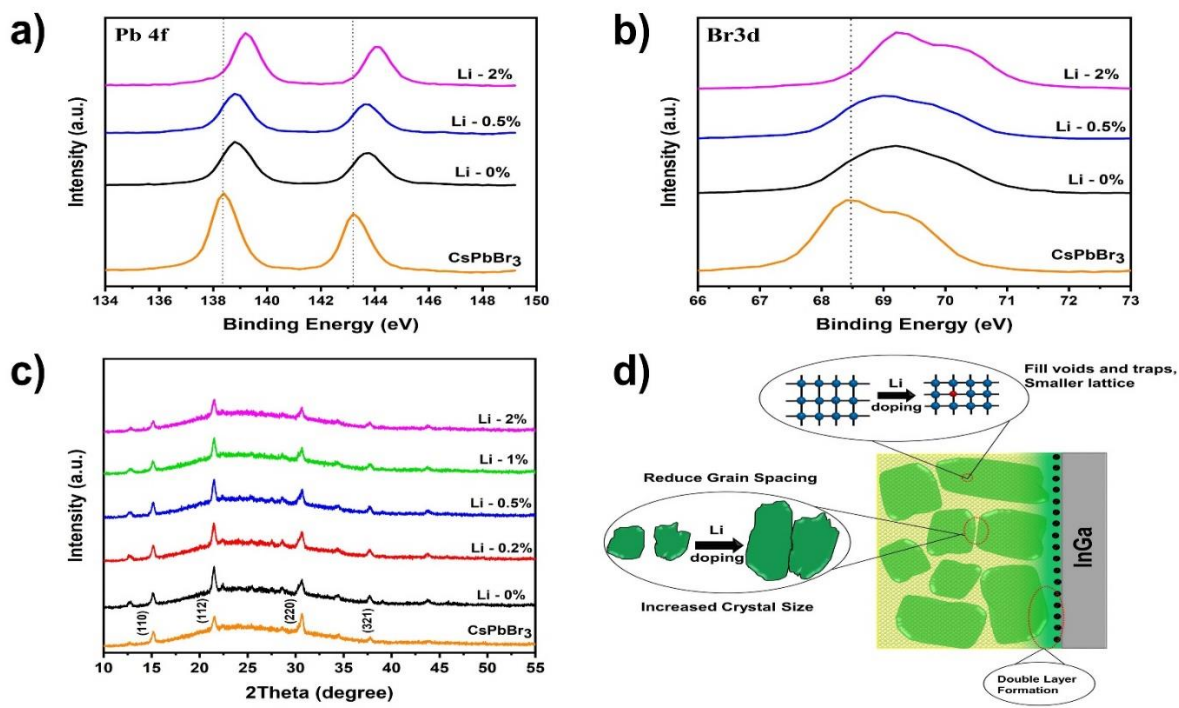

Figure 4. High resolution X-ray photoelectron spectroscopy (XPS) and X-ray diffraction (XRD) of CsPbBr$_3$:PEO:LiPF$_6$ films, as well as a conceptual illustration of Li$^+$-induced performance enhancement. High-resolution XPS spectral regions for a) Pb (4f$_{5/2}$, 4f$_{7/2}$) and b) Br (3d$_{3/2}$, 3d$_{5/2}$) peaks from CsPbBr$_3$ and CsPbBr$_3$:PEO:LiPF$_6$ thin films of various LiPF$_6$ weight ratios. c) XRD pattern of CsPbBr$_3$ and CsPbBr$_3$:PEO:LiPF$_6$ thin films of various LiPF$_6$ weight ratios. d) Conceptual illustration of the benefits afforded by LiPF$_6$ doping in PeLECs.

X-ray photoelectron spectroscopy (XPS) analysis of the high resolution spectra involving Pb 4f, Br 3d, C 1s, and F 1s (Figure 4a-b, and Supporting Information Figure S10) was conducted to further understand the effect of the LiPF$_6$ dopant in CsPbBr$_3$:PEO films. As shown in Figure 4a, the Pb 4f spectrum for the pristine CsPbBr$_3$ film has two peaks associated with 4f$_{5/2}$ and 4f$_{7/2}$ orbitals located at 138.40 and 143.20 eV, respectively, which correspond to Pb$^{2+}$ cations. The two Br 3d signals, corresponding to 3d$_{3/2}$ and 3d$_{5/2}$ peaks, are also clearly seen at 68.4 and 69.2 eV.

Addition of PEO to the CsPbBr$_3$ film shifts all of these peaks to higher binding energies, which is attributed to the interaction between PEO and metallic/ion Pb reported previously.[28] Furthermore, we also observe that adding LiPF$_6$ dopant to the CsPbBr$_3$:PEO film further shifts these peaks. These spectral shifts are a clear indication of a lattice contraction of perovskite after Li$^+$ incorporation, shorter Pb-Br bonds and higher binding energy for Pb 4f and Br 3d.[20, 29] The presence of PEO and LiPF$_6$ in films has been further confirmed from the C 1s and F 1s XPS spectra (Supporting Information Figure S10a-b). The adventitious C 1s signal is obvious at 285.2 eV for the pristine CsPbBr$_3$ thin film. However, after PEO addition, a new C 1s peak appears at 287 eV, which corresponds to the C-O-C groups of the PEO.[30] An F 1s spectral peak at 686.6 eV appears after adding LiPF$_6$ to CsPbBr$_3$:PEO, confirming the presence of the fluoride ion in the film. Also, we note that the photoemission of Li 1s could not be detected due to the very low sensitivity factor of Li. The detailed chemical bonding of the composite perovskite films was analyzed by deconvoluting the XPS peaks (Supporting Information Figure S10).

Figure 4c illustrates the X-ray diffraction (XRD) spectra of CsPbBr$_3$, CsPbBr$_3$:PEO, and CsPbBr$_3$:PEO:LiPF$_6$. The thin film XRD spectra indicate the primary diffraction peaks at 15.21°, 21.46°, and 30.70° that corresponds to diffraction planes of (110), (112) and (220), respectively, in agreement with previous reports for an orthorhombic (Pnma) crystal structure.[31-32] The peaks at 12.7° and 22.41° indicate the presence of a trigonal phase of Cs$_4$PbBr$_6$ with the diffraction planes of (102) and (213), respectively, as seen in literature, indicating a mixed phase of CsPbBr$_3$ and Cs$_4$PbBr$_6$.[33-35] In addition, we analyzed the peak widths and positions, finding successive contraction of the perovskite crystal lattice parameters with increasing amounts of LiPF$_6$ (see Table S2), suggesting Li substitution in the perovskite lattice.

Overall, these observations suggest several benefits of lithium salt addition that led to high performance (14730 cd/m$^2$, 22.4 cd/A) among single layer PeLECs that are comparable to the most efficient PeLEDs, which are typically multilayer devices (Supporting Information Table S3). These benefits are summarized in Figure 4d. First, an overall consideration of the system and a detailed view of the device performance suggests that an optimal concentration of LiPF$_6$ facilitates double layer formation, thus improving charge carrier injection. Second, SEM and AFM studies demonstrated that an optimal lithium concentration reduces the spacing between grain boundaries, leading to smooth, pinhole free films. This limits leakage current, enables high electron and hole conductivities for enhanced carrier transport. PL study correlated with XPS and XRD studies show that lithium reduces the crystal lattice parameters through filling traps and voids, reducing sources of nonradiative losses for highly efficient light emission in thin films and devices. A detailed balance of all of these features thus leads to high performance LECs of simple, single-layer architectures for next generation optoelectronic applications.

ASSOCIATED CONTENT

Supporting information available: experimental methods, electroluminescence spectra, luminance versus voltage and time with associated data, illustrations of PL mechanisms, SEM and AFM images, XPS spectra, and XRD data.

ACKNOWLEDGMENT

J.D.S. acknowledges support from the National Science Foundation (ECCS 1906505). Q.G. acknowledges support from the Welch Foundation (AT-1992-20190330). A.Z. acknowledges support from the Welch Foundation (AT-1617) and from the Ministry of Education and Science of the Russian Federation (14.Y26.31.0010).

# Supporting Information

# Bright and Efficient Perovskite Light Emitting Electrochemical Cells Leveraging Ionic Additives

*Masoud Alahbakhshi[1], Aditya Mishra[2], Ross Haroldson[3], Arthur Ishteev[6], Jiyoung Moon[2], Qing Gu[1], Jason D. Slinker[2,3]\* and Anvar A. Zakhidov[3,4,5]\**

[1]Department of Electrical and Computer Engineering, The University of Texas at Dallas, 800 West Campbell Road, Richardson, Texas 75080-3021, United States.

[2]Department of Materials Science and Engineering, The University of Texas at Dallas, 800 West Campbell Road, Richardson, Texas 75080-3021, United States.

[3]Department of Physics, The University of Texas at Dallas, 800 West Campbell Road, Richardson, Texas 75080-3021, United States.

[4]NanoTech Institute, The University of Texas at Dallas, 800 West Campbell Road, Richardson, Texas 75080-3021, United States.

[5]Department of Nanophotonics and Metamaterials, ITMO University, St. Petersburg, Moscow, Russia.

[6]Laboratory of Advanced Solar Energy, NUST MISiS, Moscow, 119049, Russia

Experimental Methods

Figures and Tables

**Experimental Methods**

**Materials:** Lead (II) bromide ($PbBr_2$; 99.99% trace metal basis), Cesium bromide (CsBr; 99.99%) and Polyethylene Oxide (PEO; M.W. > 5,000,000) were all purchased from Alfa Aesar. Lithium Hexafluorophosphate ($LiPF_6$; 99.99%) and Dimethyl Sulfoxide (DMSO; anhydrous > 99.9 %) were purchased from Sigma Aldrich. Gallium Indium eutectic (GaIn; 99.99%) was purchased from Beantown Chemical.

**Perovskite Solution Preparation:** The $CsPbBr_3$-based precursor solution was prepared by dissolving $PbBr_2$ and CsBr in a 1:1.5 molar ratio with PEO (10 mg/ml) in anhydrous DMSO solution. Then the perovskite precursor solution was stirred at 60 °C for dissolution overnight. When all solutions were dissolved, an empty vial was weighed and the desirable amount of PEO was added, then the weight difference before and after the PEO addition was measured to get an accurate weight of the viscous solution. The weight ratio of $CsPbBr_3$ to PEO was 100:80. Finally, these solutions were blended with 4 mg/ml DMSO solutions of $LiPF_6$ to prepare mixtures of five different weight ratios (0.2%, 0.5%, 1%, 2% and 5%) of lithium salt with the perovskite-polymer composition.

**Device Fabrication:** The ITO/glass substrates (Liasion Quartz (Lianyungang Jiangsu China), sheet resistance ~ 15 Ω sq$^{-1}$) were cleaned sequentially with detergent solution, deionized water, acetone, Toluene and 2-propanol in an ultra-sonication bath for 15 mins. Subsequently, the substrates were dried with nitrogen and treated for 20 min with UV-ozone. To obtain $CsPbBr_3$:PEO:$LiPF_6$ thin films, precursor solutions were spin-coated onto ITO substrates at 1200 rpm for 45 min. Then, the thin film was put under vacuum for 1 minute to have a uniform and pinhole-free thin film. Finally, the $CsPbBr_3$:PEO:$LiPF_6$ film was annealed at 150 °C for 15 seconds to remove the residual solvent.

**Electroluminescence Measurements:** All measurements were conducted using a mechanical probe-station under high vacuum <10mTorr. Current density-voltage (J-V) and luminance-voltage (L-V) characteristics were measured using a Keithley 2400 source meter and a Photo Research PR-650 spectroradiometer in the range of 0V to 6V with a 0.3V increment.

**Photoluminescence *vs* Time Measurements:** Measurements were taken using an Ocean Optics QE65000 spectrometer coupled with a fiber optic cable pointed at the thin films through a 450nm longpass dielectric filter to block out the 405nm CW laser diode excitation signal.

**Photoluminescence Quantum Efficiency Measurements:** Thin film samples deposited on glass were attached to a custom holder then placed inside a Spectral Physics integrating sphere. Samples were excited at a 15 degree angle incidence to avoid back reflecting excitation light out of the integrating sphere. Using the method developed by de Mello *et al*., we took measurements of the excited thin films both in and out of the excitation beam path [1]. Fluctuations in excitation power were monitored by splitting the beam with a beam splitter and placing a photodiode power meter hooked up to a Thorlabs PDA200C photodiode amplifier. The collimated beam profile was shaped with a precision cut 1000 micron diameter circular aperture. The power density was then measured to be 561±4 mW/cm$^2$. A fiber optic cable was mounted to one of the integrating sphere ports and was coupled to the Ocean Optics QE65000 spectrometer where spectrum measurements were taken.

**Scanning Electron Microscopy (SEM):** Secondary electron SEM images were taken with a Zeiss Supra-40 SEM using an in-lens detector at an accelerating voltage of 10kV.

**Atomic Force Microscopy (AFM):** The AFM images were performed using a Veeco Model 3100 Dimension V to scrutinize the morphology of thin films. The thin films were scanned for 5μm ×

5μm area at 0.8Hz rate using an OTESPA-R3 AFM tip from Bruker. Tapping mode AFM was used for this characterization.

**X-ray Photoelectron Spectroscopy (XPS):** XPS measurements were conducted on perovskite thin films using a Versa Probe II at an ultrahigh vacuum of $10^{-8}$ Pa. The X-ray source was an Al Kα which has 1486.6eV photon energy, 50W gun power, 15kV operating voltage, 200um X-ray spot size, and 59° angle between the X-ray source and detector. The calibration using internal standard Au, Cu, and Ag samples was performed before obtaining the XPS data. Data was collected without injecting a flux of low energy electrons.

**X-Ray Diffraction (XRD):** XRD measurements were collected using a Rigaku SmartLab X-ray Cu target (Ka1=1.54059 Å) and a HyPix 3000 detector. The 2-theta/omega scan was consistently performed in the 2-theta range of 10° to 55° with a 0.01° step and a ~1°/min scan speed.

**Supporting Information Figure S1**

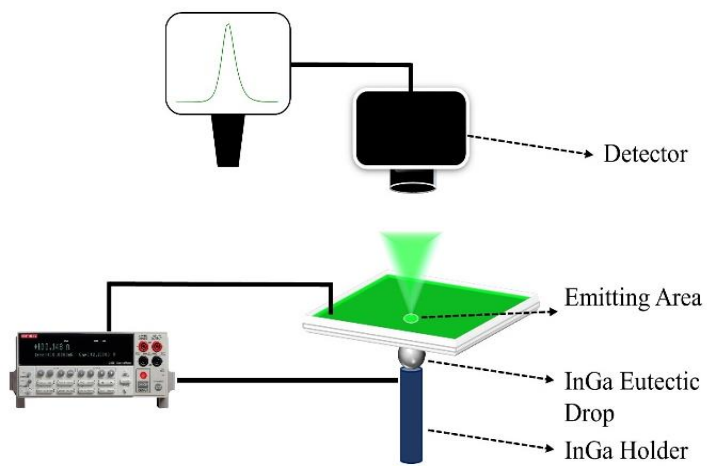

Figure S1. The experimental setup for capturing the electroluminescence performance of In-Ga/CsPbBr$_3$:PEO:LiPF$_6$/ITO LEC devices.

**Supporting Information Figure S2**

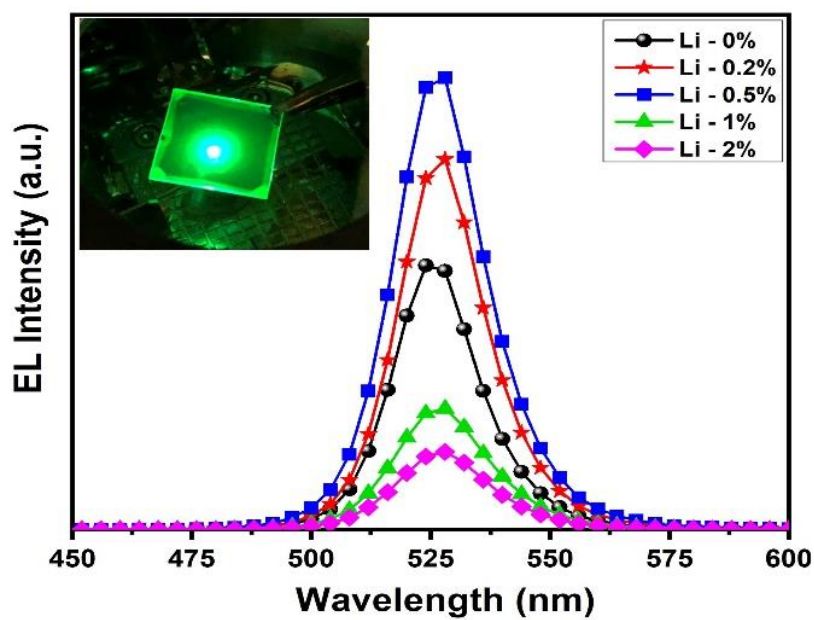

Figure S2. Electroluminescence spectra of In-Ga/CsPbBr$_3$:PEO:LiPF$_6$/ITO light emitting electrochemical cells with various weight ratios of LiPF$_6$. The inset shows the electroluminescence of a 0.5% LiPF$_6$ PeLEC operating at 5 volts.

**Supporting Information Table S1**

| Devices | Turn-on Voltage (V) | Max CE (cd/A) | Max Lum (cd/m$^2$) | Max Current Density (mA/cm$^2$) |
|---|---|---|---|---|
| CsPbBr$_3$:PEO (0%) | 2.5 | 9.0 | 8175 | 2390 |
| CsPbBr$_3$:PEO:LiPF$_6$ (0.2%) | 2.2 | 17.2 | 11800 | 2952 |
| CsPbBr$_3$:PEO:LiPF$_6$ (0.5%) | 1.9 | 22.4 | 14730 | 3301 |
| CsPbBr$_3$:PEO:LiPF$_6$ (1%) | 2.4 | 4.5 | 5675 | 2150 |
| CsPbBr$_3$:PEO:LiPF$_6$ (2%) | 2.4 | 4.1 | 2619 | 1486 |
| CsPbBr$_3$:PEO:LiPF$_6$ (5%) | 2.7 | 1.2 | 517 | 749 |

Table S1. Summrized performance of In-Ga/CsPbBr$_3$:PEO:LiPF$_6$/ITO LEC devices with different ratios of LiPF$_6$.

**Supporting Information Figure S3**

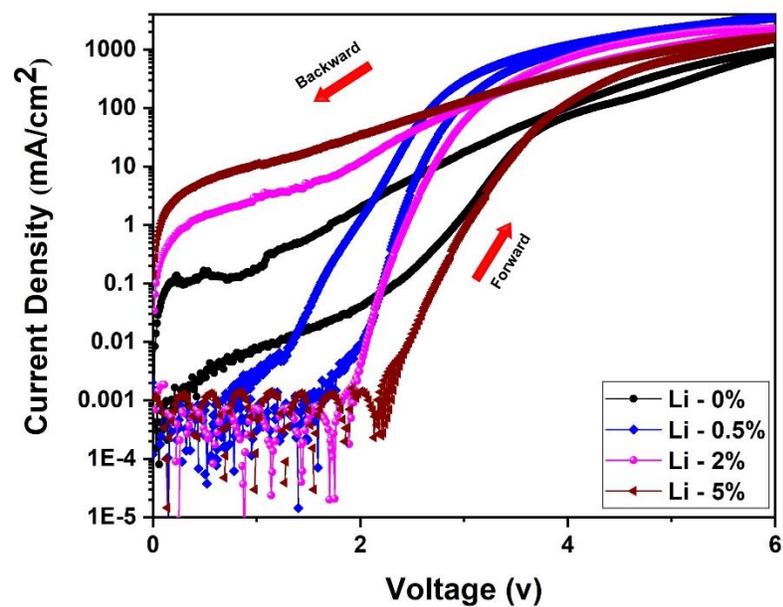

Figure S3. Current density *versus* voltage of In-Ga/CsPbBr$_3$:PEO:LiPF$_6$/ITO LEC devices with different ratios of LiPF$_6$. The graph clearly shows that in the optimized ratio of Li (0.5%), the hysteresis state decreases significantly.

**Supporting Information Figure S4**

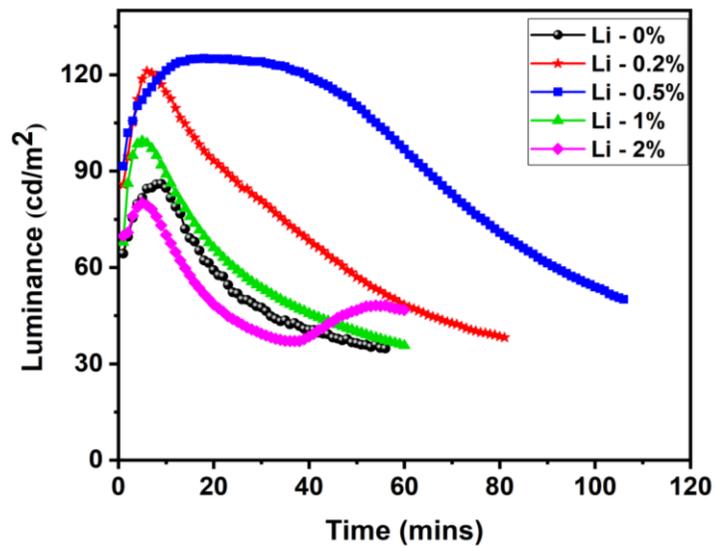

Figure S4. Luminance versus time for constant voltage operation of In-Ga/CsPbBr$_3$:PEO:LiPF$_6$/ITO LEC devices with different ratios of LiPF$_6$, demonstrating increased stability with optimal LiPF$_6$ concentration.

**Supporting Information Figure S5**

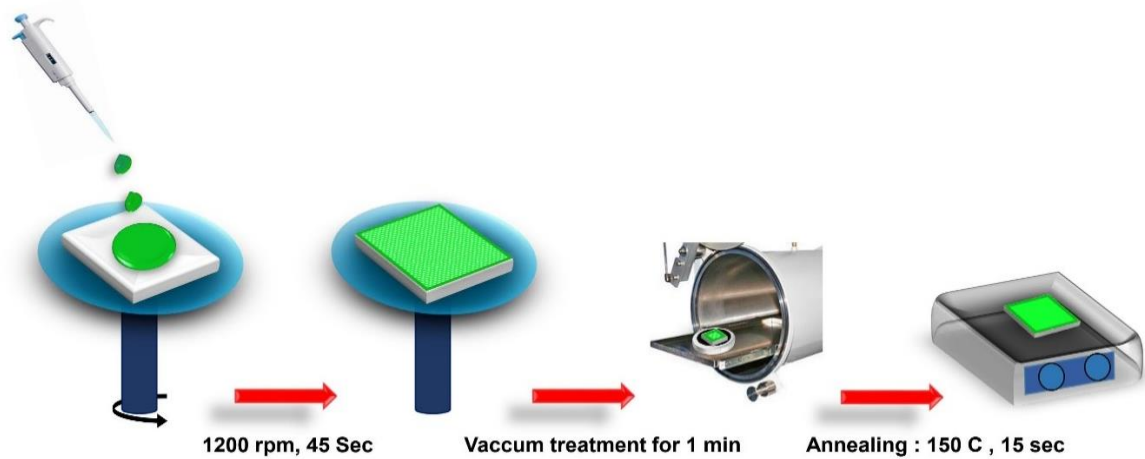

Figure S5. Illustration of the fabrication of thin films of $CsPbBr_3$:PEO:$LiPF_6$. Initially, components are spin cast at 1200 rpm for 45 seconds, followed by rapid vacuum treatment for 1 minute and annealing for 15 seconds at 115 °C.

**Supporting Information Figure S6**

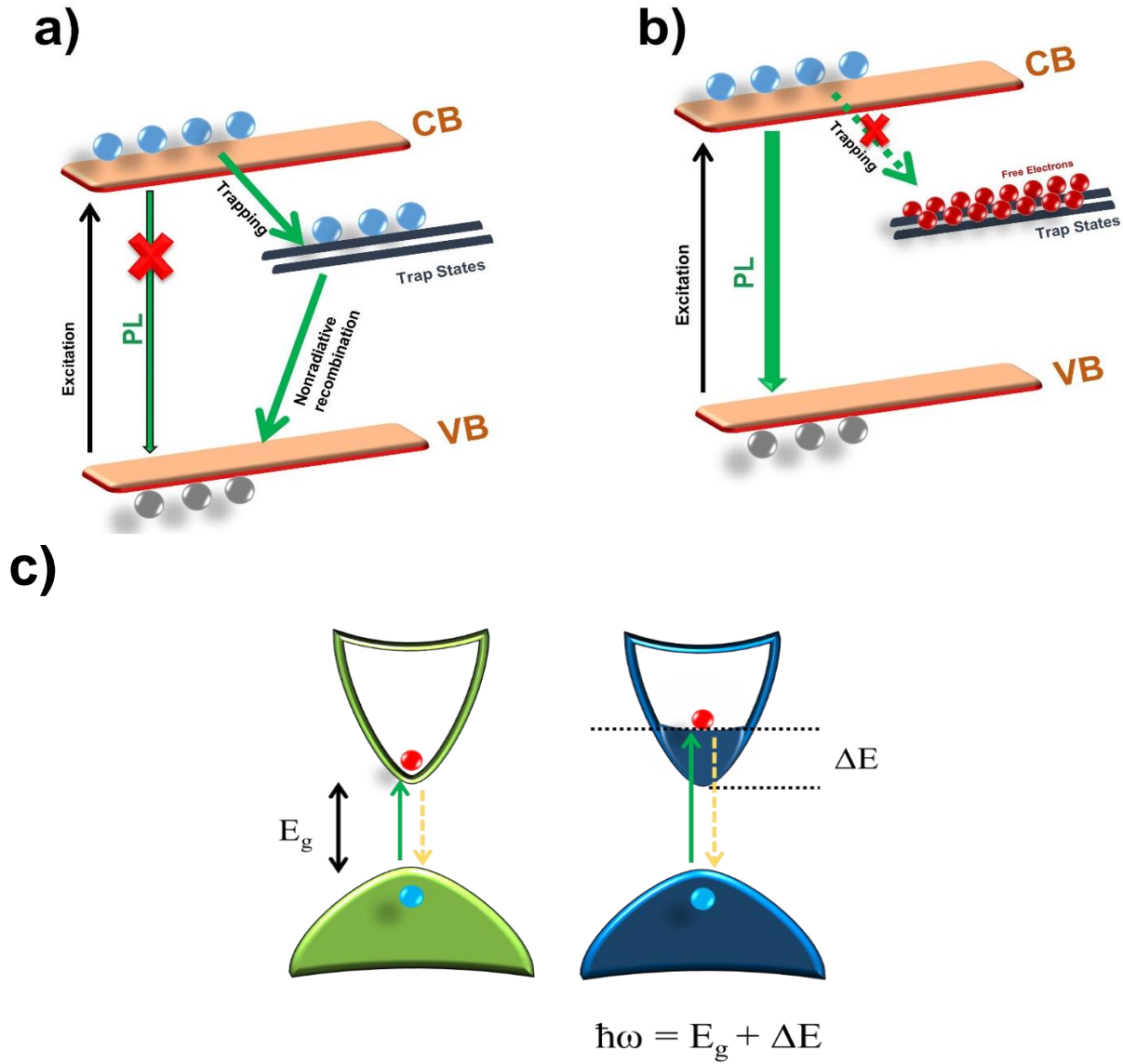

Figure S6. The mechanism of the recombination of electrons and holes for undoped thin films (a) for doped thin films (b) by Li. The doping of Li produces free electrons to fill and passivate the trap states, therefore the radiative recombination will improve. c) Illustration of the Burstein–Moss effect causing a PL blue shift.

**Supporting Information Table S2**

| Sample | Average grain size (nm) | FWHM (degree) | 2θ (degree) | Crystallite size (nm) |
|---|---|---|---|---|
| CsPbBr$_3$ | 145 | 0.51 | 21.53 | 16.49 |
| | | 0.50 | 30.52 | 17.04 |
| CsPbBr$_3$:PEO (0%) | 79 | 0.52 | 21.5 | 16.39 |
| | | 0.53 | 30.55 | 16.15 |
| CsPbBr$_3$:PEO:LiPF$_6$ (0.2%) | 105 | 0.51 | 21.49 | 16.47 |
| | | 0.54 | 30.59 | 16.07 |
| CsPbBr3:PEO:LiPF$_6$ (0.5%) | 163 | 0.53 | 21.49 | 16.04 |
| | | 0.57 | 30.57 | 15.15 |
| CsPbBr$_3$:PEO:LiPF$_6$ (1%) | 166 | 0.58 | 21.52 | 14.56 |
| | | 0.58 | 30.58 | 14.79 |
| CsPbBr$_3$:PEO:LiPF$_6$ (2%) | 150 | 0.61 | 21.55 | 13.91 |
| | | 0.63 | 30.65 | 13.71 |
| CsPbBr$_3$:PEO:LiPF$_6$ (5%) | 161 | 0.65 | 21.6 | 13.04 |
| | | 0.65 | 30.7 | 13.20 |

Table S2. Average grain size as measured from SEM and XRD spectral data for CsPbBr$_3$, CsPbBr$_3$:PEO, and CsPbBr$_3$:PEO:LiPF$_6$ thin films.

**Supporting Information Figure S7**

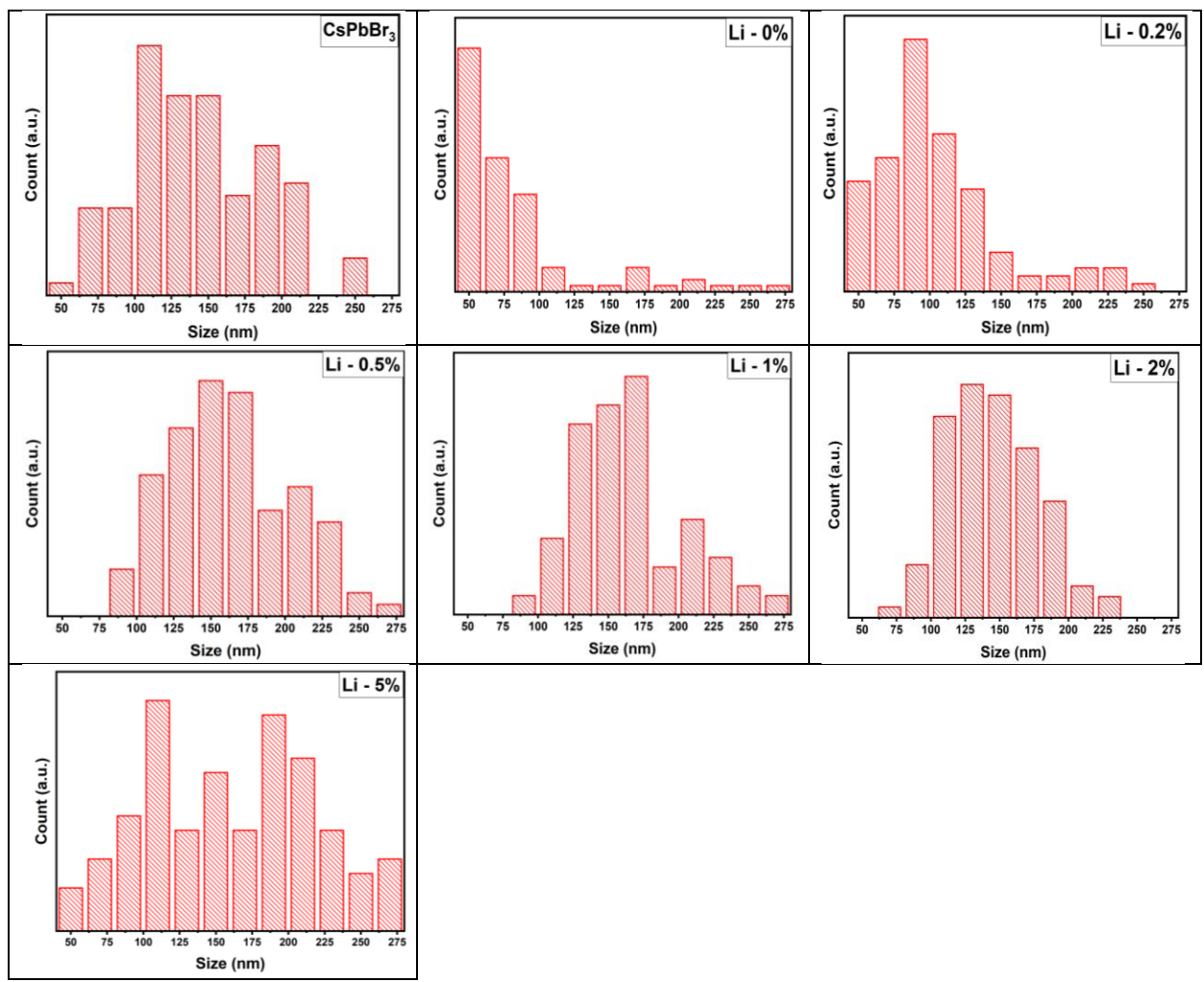

Figure S7. Histograms of $CsPbBr_3$, $CsPbBr_3$:PEO, and $CsPbBr_3$:PEO:$LiPF_6$ thin film grain size as measured from SEM imaging.

**Supporting Information Figure S8**

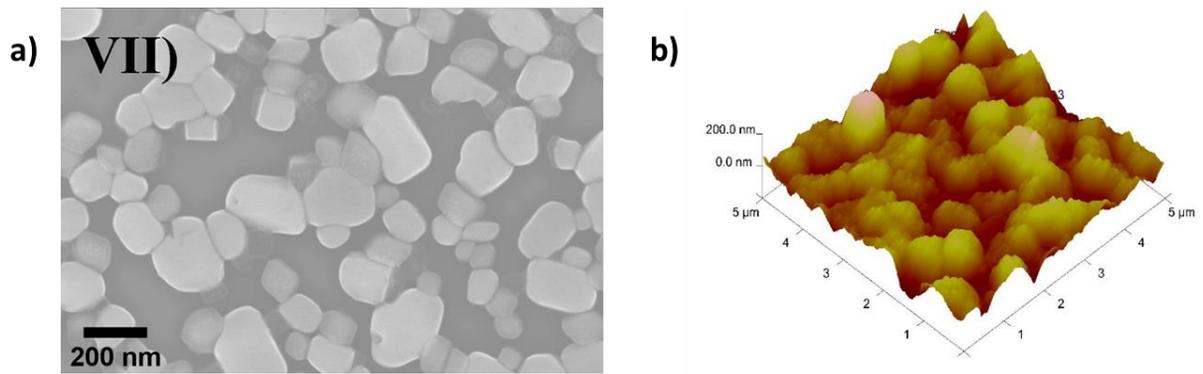

Figure S8. Morphological impact of high LiPF$_6$ concentrations in CsPbBr$_3$:PEO:LiPF$_6$ thin films. a) SEM and b) AFM images of CsPbBr$_3$:PEO:LiPF$_6$ (5%).

**Supporting Information Figure S9**

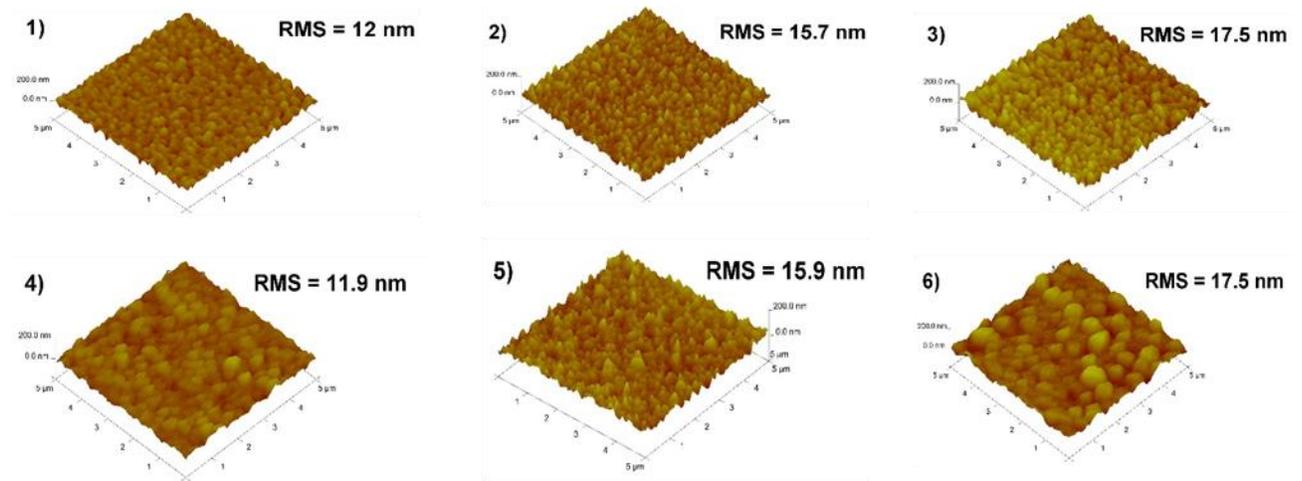

Figure S9. 3D AFM topography images of 1) CsPbBr$_3$, 2) CsPbBr$_3$:PEO (0%), 3) CsPbBr$_3$:PEO:LiPF$_6$ (0.2%), 4) CsPbBr$_3$:PEO:LiPF$_6$ (0.5%), 5) CsPbBr$_3$:PEO:LiPF$_6$ (1%), 6) CsPbBr$_3$:PEO:LiPF$_6$ (2%).

## Supporting Information Figure S10

Figure 10. X-Ray photoelectron spectroscopy (XPS) spectra for CsPbBr$_3$, CsPbBr$_3$:PEO, CsPbBr$_3$:PEO:LiPF$_6$.

**CsPbBr$_3$**

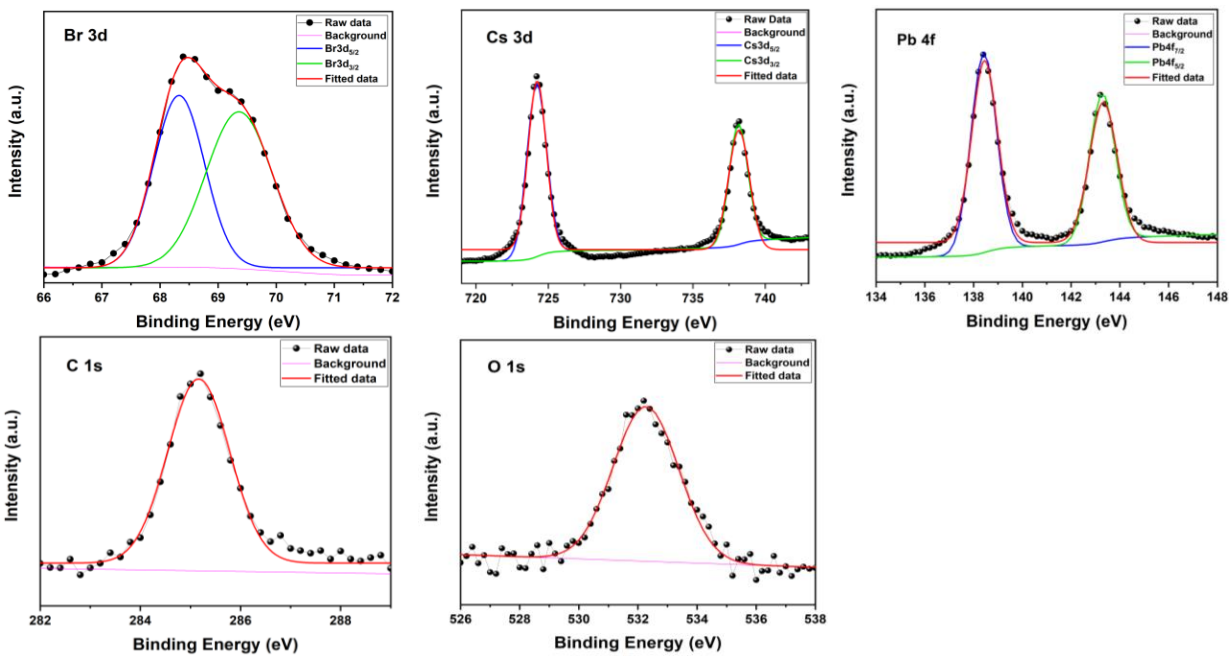

**CsPbBr$_3$:PEO**

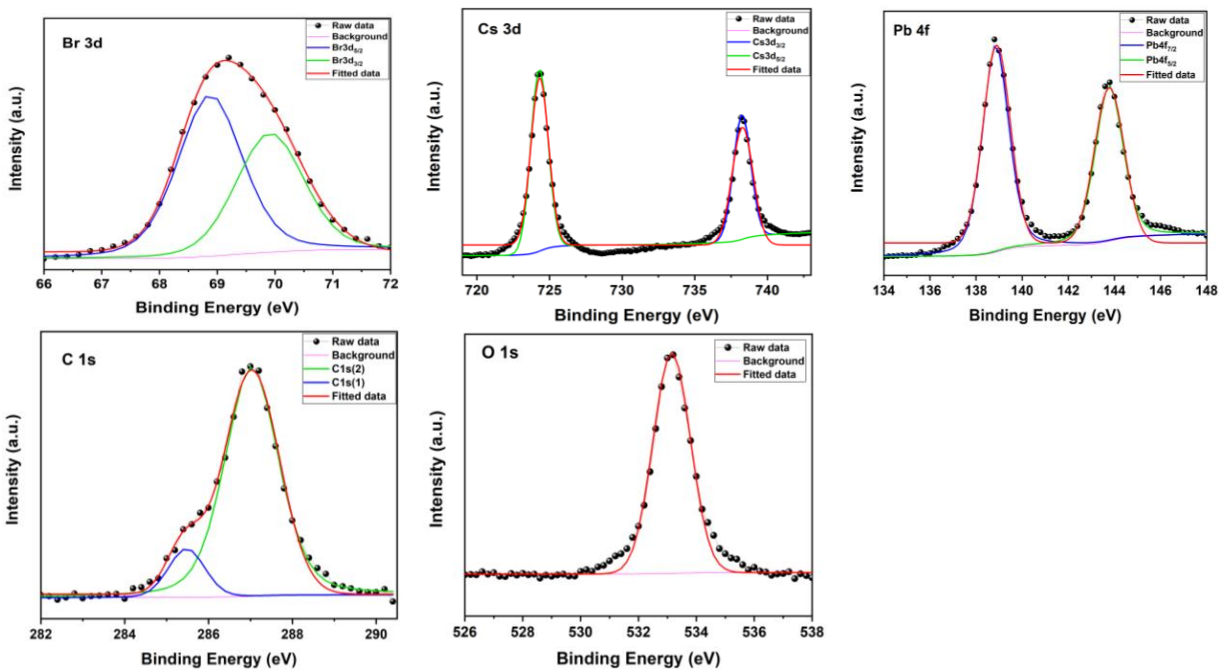

**CsPbBr$_3$:PEO:LiPF$_6$ (0.5%)**

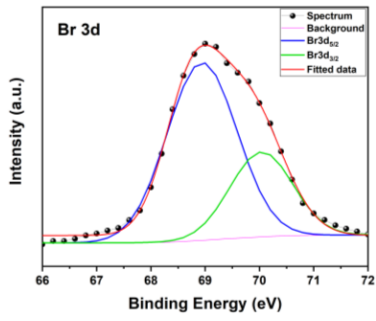
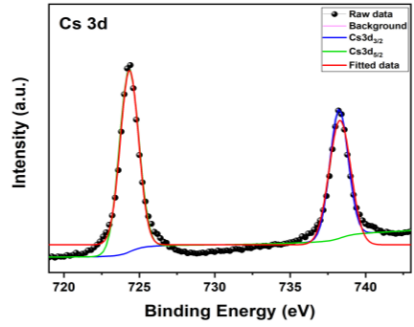
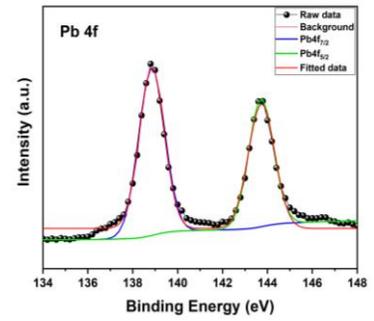
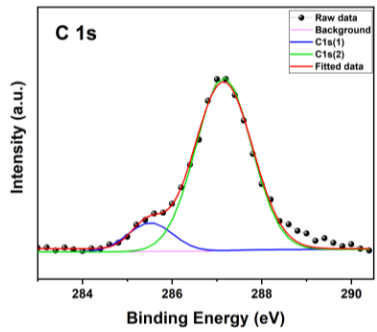
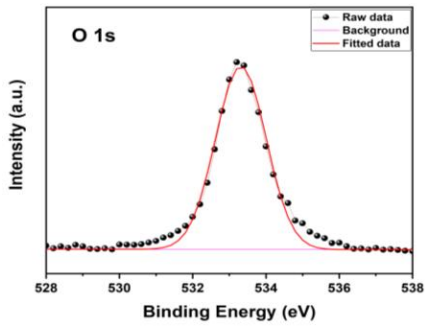
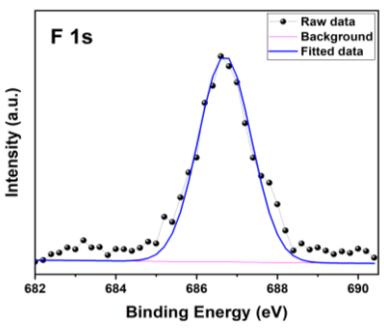

**CsPbBr$_3$:PEO:LiPF$_6$ (2%)**

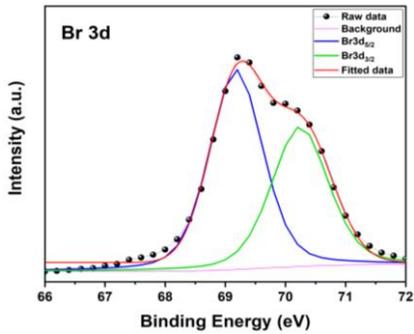
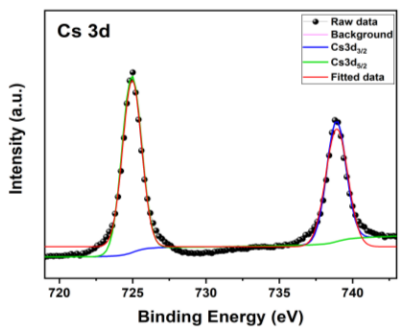
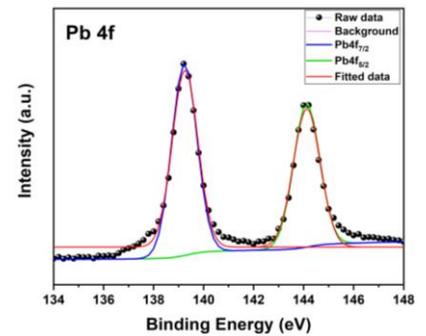
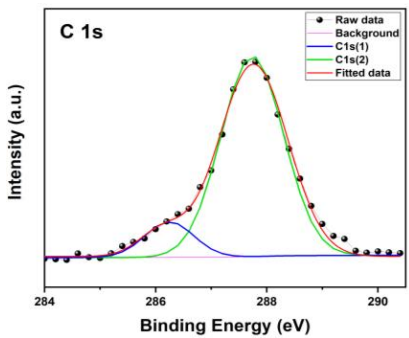
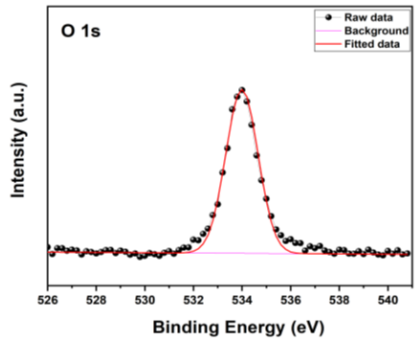
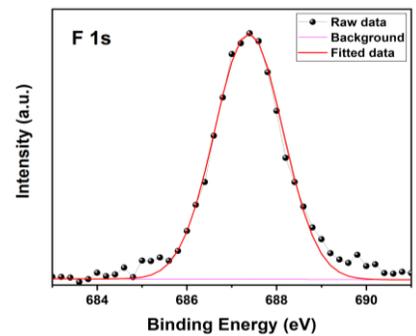

**Supporting Information Table S3**

| Device structure | Single or multilayer | Light Emitting Mechanism | Turn on voltage | Luminescence Max (cd/m2) | Current Efficiency (cd/A) | Ref |
|---|---|---|---|---|---|---|
| ITO / PEDOT:PSS /PKNP:TMPE:LiCF3SO3/Al | M | LEC | 11 | 1.8 | 0.013 | [2] |
| ITO / Pero-PEO-composite / In (Ga,Au) | S | LED | 3 | 4064 | 0.74 | [3] |
| ITO / Pero-PEO-PVP-composite / InGa | S | LED | 1.9 | 593178 | 21.5 | [4] |
| ITO / PEDOT:PSS /Perovskite / MoO3/Au | M | LEC | 2.4 | 157 (0.23 W/ Sr. m$^2$) | - | [5] |
| ITO / Pero-PEO composite / Ag NW | S | LED | 2.6 | 21014 | 4.9 | [6] |
| ITO / PEDOT:PSS /MAPbBr3 / SPB-02T /LiF/ Ag | M | LED | 2 | 3490 | 0.43 | [7] |
| ITO/PEDOT:PSS/CsPbBr3:PEO/TPBI/LiF/Al | M | LED | 2.6 | 51890 | 21.38 | [8] |
| ITO/PEDOT:PSS/CsPbBr3:PEG/TPBI/LiF/Al | M | LED | 2.5 | 36600 | 19 | [9] |
| ITO / PEDOT:PSS / CsPbBr$_{1.25}$I$_{1.75}$ NC: KCF3SO3/Al | M | LEC | 4 | 8 | - | [10] |
| VACNT / SC-MAPbBr3 / VACNT | S | LEC | 26 | 1000 | - | [11] |
| **Our Work** | S | LEC | 1.9 | 14730 | 22.4 | |

Table S3. Comparative performances of the best-in-class perovskite based LEDs and LECs.